\documentclass{mn2e}
\usepackage{epsfig}

\title[Mid-IR imaging of Source G in W49A]{Massive star formation and
  feedback in W49A: The source of our Galaxy's most luminous water
  maser outflow\thanks{Based on observations obtained at the Gemini
    Observatory, which is operated by the Association of Universities
    for Research in Astronomy (AURA) under a cooperative agreement
    with the NSF on behalf of the Gemini partnership: the National
    Science Foundation (United States), the Science and Technology
    Facilities Council (United Kingdom), the National Research Council
    (Canada), CONICYT (Chile), the Australian Research Council
    (Australia), CNPq (Brazil) and CONICET (Argentina).}}

\author[N.\ Smith et al.]{ Nathan Smith$^1$\thanks{Email:
    nathans@astro.berkeley.edu}, Barbara A.\ Whitney$^2$, Peter S.\
  Conti$^3$, Chris G.\ De~Pree$^4$, \newauthor and James M.\ Jackson$^5$  \\
  $^1$Astronomy Department, University of California, 601 Campbell
  Hall, Berkeley, CA 94720, USA \\ $^2$ Space Science Institute, 3100
  Marine Street, Suite A353, Boulder, CO 80303, USA \\ $^3$ JILA,
  Campus Box 440, University of Colorado, Boulder, CO 80309, USA \\
  $^4$ Department of Physics and Astronomy, Agnes Scott College, 141
  East College Avenue, Decatur, GA 30030, USA \\ $^5$ Institute for
  Astrophysical Research, Boston University, 725 Commonwealth Ave.,
  Boston, MA 02215, USA}

\begin{document}
\date{Accepted 0000, Received 0000, in original form 0000}
\pagerange{\pageref{firstpage}--\pageref{lastpage}} \pubyear{2002}
\def\arcdeg{\degr}
\maketitle
\label{firstpage}

\begin{abstract}

  We present high spatial resolution mid-infrared (IR) images of the
  ring of ultracompact H~{\sc ii} regions in W49A obtained at Gemini
  North, allowing us to identify the driving source of its powerful
  H$_2$O maser outflow.  These data also confirm our previous report
  that several radio sources in the ring are undetected in the mid-IR
  because they are embedded deep inside the cloud core.  We locate the
  source of the water maser outflow at the position of the compact
  mid-IR peak of source G (source G:IRS1) to within 0$\farcs$07.  This
  IR source is not coincident with any identified compact radio
  continuum source, but is coincident with a hot molecular core, so we
  propose that G:IRS1 is a hot core driving an outflow analogous to
  the wide-angle bipolar outflow in OMC-1.  G:IRS1 is at the origin of
  a larger bipolar cavity and CO outflow.  The water maser outflow is
  orthogonal to the bipolar CO cavity, so the masers probably reside
  near its waist in the thin cavity walls.  Models of the IR emission
  require a massive protostar with $M_*\simeq$45 $M_{\odot}$,
  $L_*\simeq$3$\times$10$^5$ L$_{\odot}$, and an effective envelope
  accretion rate of $\sim$10$^{-3}$ $M_{\odot}$ yr$^{-1}$.
  Feedback from the central star could potentially drive the
  small-scale H$_2$O maser outflow, but it has insufficient radiative
  momentum to have driven the large-scale bipolar CO outflow,
  requiring that this massive star had an active accretion disk over
  the past 10$^4$ yr.  Combined with the spatialy resolved morphology
  in IR images, G:IRS1 in W49 provides compelling evidence for a
  massive protostar that formed by accreting from a disk, accompanied
  by a bipolar outflow.

\end{abstract}

\begin{keywords}
  H~{\sc ii} regions --- ISM: individual (W49A) --- ISM: jets and
  outflows --- stars: formation --- stars: pre--main-sequence
\end{keywords}

\section{INTRODUCTION}

Because of their extremely short Kelvin-Helmholtz timescales, massive
protostars above 8 $M_{\odot}$ begin burning H while still accreting
and while still buried deep in their natal cloud cores.  When the
massive star reaches the main-sequence and ionizes its surroundings,
it forms an ultracompact H~{\sc ii} (UCHII) region (Wood \& Churchwell
1989; Churchwell 2002; Hoare et al.\ 2007), with typical sizes
$\la$0.1~pc.  Precursors of UCHII regions are thought to be dense hot
molecular cores (e.g., Kurtz et al.\ 2000), but early accretion phases
are not as well understood as they are for low-mass stars.  With
spherical symmetry, radiation pressure on dust would lead to a limit
at about 10~$M_{\odot}$ for the mass that could be accreted onto a
star (e.g., Wolfire \& Cassinelli 1987; Kahn 1974; Larson 1969), or
somewhat higher if momentum of infalling matter can overwhelm the dust
sublimation radius.  The way in which massive stars overcome this
difficulty to reach $\sim$40-150~$M_{\odot}$ is still a topic of
current research.

One proposed solution is geometric: in scaling up the traditional view
of low-mass star formation that involves accretion disks and
collimated outflows (Keto 2003; McKee \& Tan 2003), a massive
protostar may circumvent the radiation pressure problem with
non-spherical geometry.  With an optically thick disk and an optically
thin bipolar flow, the disk can potentially shadow infalling material
while radiative luminosity can, in principle, escape out the polar
cavities without halting the accretion (Nakano 1989; Jijina \& Adams
1996; Yorke \& Sonnhalter 2002; Krumholz et al.\ 2005).  Recent
numerical simulations by Krumholz et al.\ (2009) argue that radiation
pressure does not halt accretion, allowing the formation of very
massive stars and multiple star systems from a common core.  On the
other hand, massive stars tend to form in dense clustered environments
where the effects of their neighbors may influence the star formation
process.  In clustered environments, coalescence of lower-mass
protostars has been suggested as another way to get past the
radiation-pressure limit (e.g., Bonnell et al.\ 1998; Bonnell \& Bate
2002; Stahler et al.\ 2000; Bally \& Zinnecker 2005).  It is not yet
clear from observations if the geometry of a disk plus a bipolar
cavity is associated with the formation of the most massive stars up
to $\sim$150 $M_{\odot}$ (Figer 2005; Kroupa 2005), while
fragmentation in massive disks may also interfere with forming the
most massive stars by accretion (Kratter \& Matzner 2006; Krumholz et
al.\ 2009).

The massive star-forming complex W49A is an excellent laboratory in
which to investigate the processes occurring during the enshrouded
early lives of massive stars, despite its large distance of 11.4~kpc
(Gwinn et al.\ 1992).\footnote{W49B is an unrelated supernova remnant
  located nearby in projection (Keohane et al.\ 2007).}  W49A contains
one of the richest known clusters of UCHII regions distributed across
only a few pc, buried inside one of the most massive ($\sim$10$^6$
$M_{\odot}$) giant molecular cloud cores in the Galaxy, with a
cumulative luminosity of well over 10$^7$ $L_{\odot}$ (see Becklin et
al.\ 1973; Dreher et al.\ 1984; Welch et al.\ 1987; Dickel \& Goss
1990; De~Pree et al.\ 1997, 2000; Smith et al.\ 2000).  Its content of
massive stars is similar to famous giant H~{\sc ii} regions like 30
Doradus, the Carina Nebula, and NGC~3603, but it is younger, with most
of its O-type stars still embedded deep in the molecular cloud.  Wood
\& Churchwell (1989) noted several different morphological types of
UCHII regions: spherical/unresolved, cometary, shell, irregular or
multiple-peaked, and bipolar (added more recently in place of
core/halo), all of which are seen in W49 (De~Pree et al.\ 1997, 2005).

This compact cluster of 40--50 UCHII regions in W49 introduced the
UCHII region ``lifetime problem'' because the sound crossing time of
this region exceeds the expected duration of the free expansion phase
for any single UCHII region (Welch et al.\ 1987).  (This
synchronization may also imply that the burst of star formation was
triggered by an external agent; Welch et al.\ 1987.)  A statistical
expression of the UCHII region lifetime problem is that the fraction
of UCHII regions compared to exposed O-type stars in the Galaxy
suggests that the UCHII region lifetime is a significant fraction of
the O star lifetime, and is therefore considerbly longer than the free
expansion time of the H~{\sc ii} region (see Wood \& Churchwell 1989).
Note, however, that Bourke et al.\ (2005) suggest that some low-mass
protostars may contaminate the Wood \& Churchwell sample, so the
implied lifetimes may be shorter (see also Hoare et al.\ 2007).

About a dozen of these UCHII regions are arranged in a remarkable 2~pc
diameter ring at the center of W49 (Welch et al.\ 1987), and the
origin of this ring remains unknown.  The brightest member of the ring
in the radio continuum is denoted source G, which breaks up into
several components at sub-arcsecond resolution (De Pree et al.\ 1997,
2000).  Source G is also the brightest source in the ring in the
thermal-infrared (IR) (Smith et al.\ 2000).

Source G is of particular interest because it harbors the most
luminous H$_2$O maser outflow known in the Milky Way (Gwinn, Moran, \&
Reid 1992).  The spatial confusion between radio continuum sources,
H$_2$O masers, and mid-IR emission has not been fully unraveled,
however.  This is, in fact, one of the main results of the present
study: we identify the driving source of the H$_2$O maser outflow by
its thermal-IR emission from dust, and it is apparently {\it not} an
ionized UCHII region.  On larger scales, mid-IR emission is
well-correlated with radio continuum emission from UCHII regions in
W49 (Smith et al.\ 2000) and in other UCHII regions (e.g.,
G29.96$-$0.2; De Buizer et al.\ 2002), but this is generally not the
case for H$_2$O maser sources, which tend to have point-like IR
emission but only weak or undetected radio emission (Tofani et al.\
1995).  High angular resolution is key, due to W49's large distance
from Earth of 11.4 kpc.  There have only been two previous mid-IR
imaging studies of source G (Becklin et al.\ 1973; Smith et al.\
2000), and one low-resolution mid-IR spectrum has been obtained
(Gillet et al.\ 1975).  None of these had sufficient angular
resolution for meaningful comparison with the complex, multi-peaked
radio continuum structure.  The mid-IR imaging with the 8~m Gemini
Observatory helps remedy this, with angular resolution comparable to
that achieved toward W49 with the VLA at centimeter wavelengths.

We present our new mid-IR Gemini images in \S 2, including a
discussion of how we spatially aligned the radio and IR data.  Then in
\S 3 we discuss results for sources in the W49 ring other than source
G, and in \S 4 we discuss source G in more detail, comparing our new
Gemini images to the radio continuum and maser emission.  This allows
us to construct a geometric working model for the complex source, and
to model the IR emission in order to constrain the physical paramters
of the embedded driving source. Finally, in \S 5 we discuss the
significance of these observations of source G and corresponding
implications for the process of massive star formation and feedback.

\begin{figure*}\begin{center}
\epsfig{file=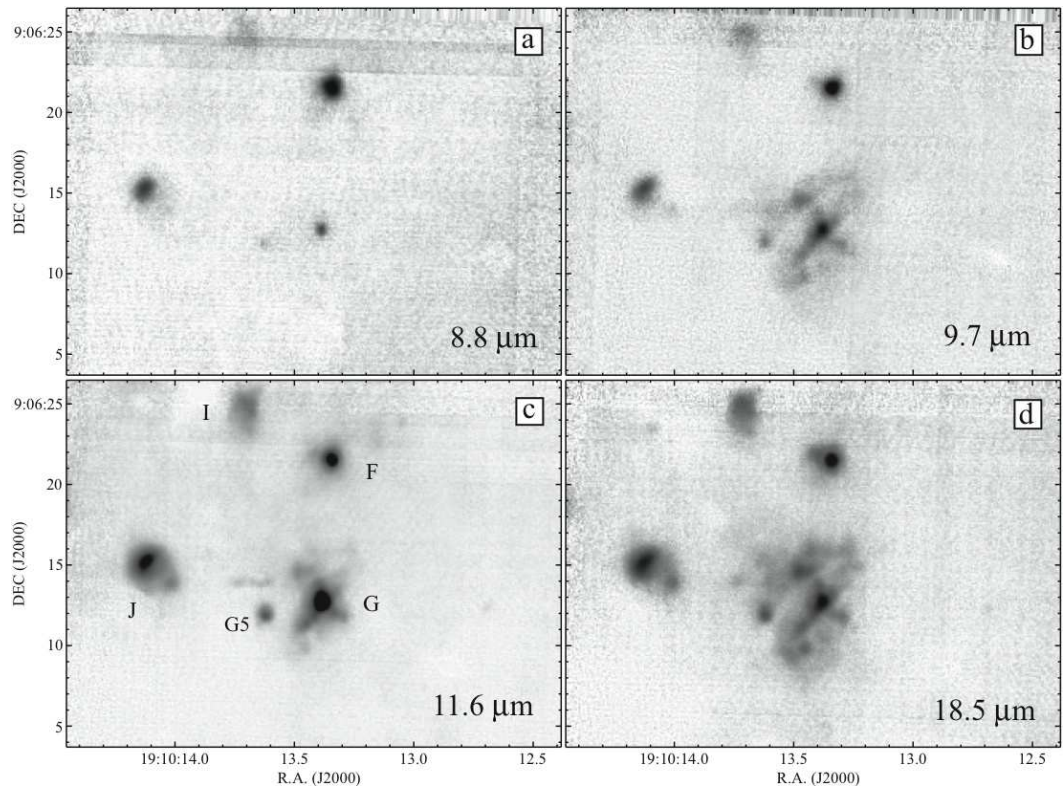,width=5.5in}
\end{center}
\caption{Grayscale representation of the Gemini North/Michelle images
  of part of the ring of UCHII regions in W49A at 8.8 $\mu$m (a), 9.7
  $\mu$m (b), 11.6 $\mu$m (c), and 18.5 $\mu$m (d). The field of view
  includes the IR sources F, G, I, and J.  The field also includes
  sources A, B, C, D, E, and H, but these are not detected in these
  images at mid-IR wavelengths.}\label{fig:one}
\end{figure*}

\begin{figure*}\begin{center}
\epsfig{file=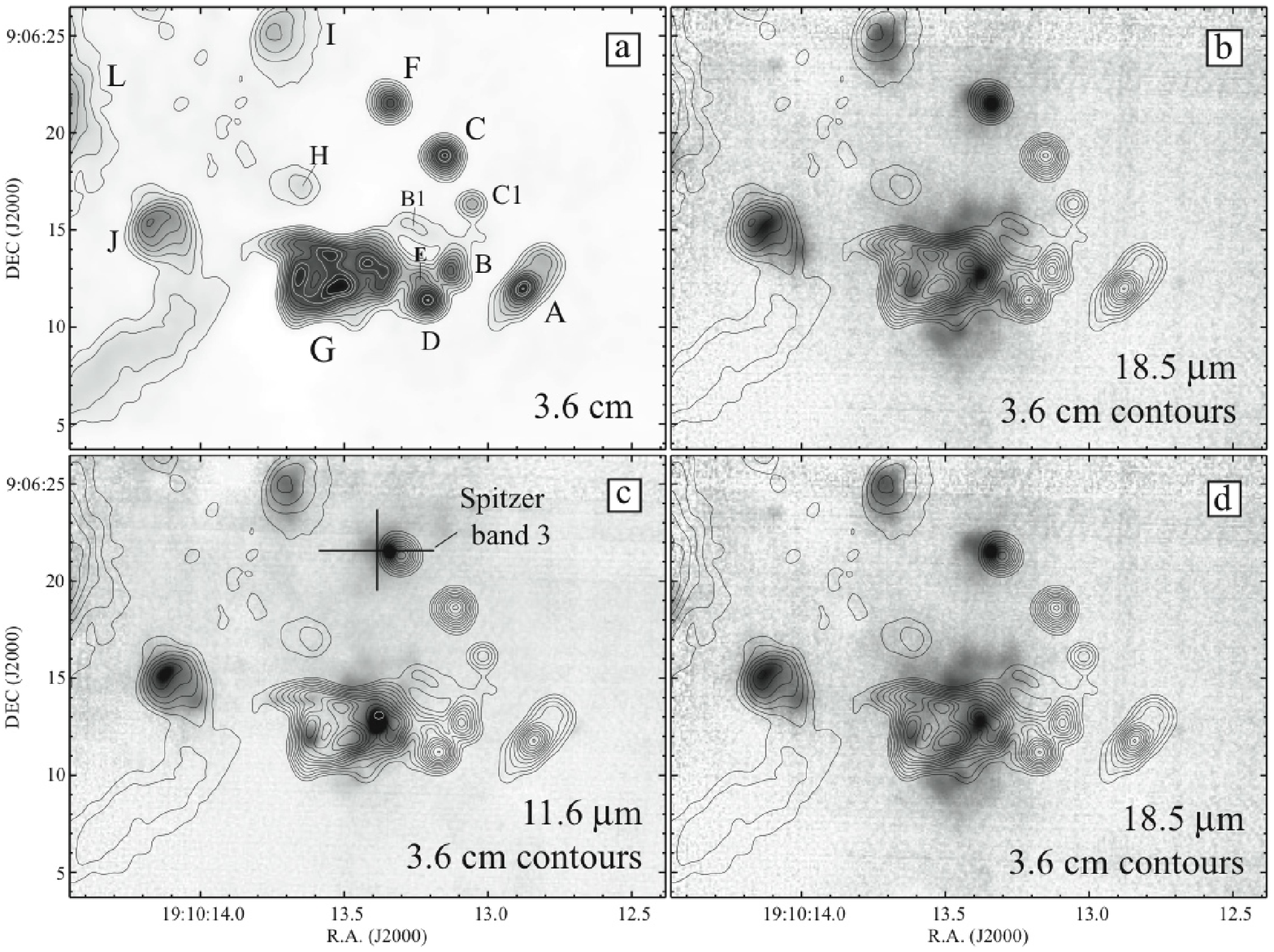,width=5.5in}
\end{center}
\caption{Comparison of IR and radio data.  (a) The 3.6~cm radio
  continuum image of the ring of UCHII regions in W49A from Depree et
  al.\ (1997) shown in grayscale and contours.  These same contours
  are superposed on mid-IR Michelle images in the remaining panels
  with two different registration options (see text \S 2.2). Panel (b)
  shows the radio contours superposed on the 18.5 $\mu$m image using
  source F as a common point for registration.  Panels (c) and (d)
  show a different spatial registration of the radio and IR images
  (the same registration is used for two different wavelengths in
  these two panels) which uses a cross-correlation of sources I and J
  for alignment.  The cross-hairs in Panel (c) show the centroid
  position of source F in Spitzer/IRAC band 3 (5.8 $\mu$m) data (the
  length of the cross hairs is for clarity; it is not meant to
  represent the much smaller positional uncertainty of
  0$\farcs$3).}\label{fig:two}
\end{figure*}

\section{OBSERVATIONS}

\subsection{Gemini North/Michelle Images}

We obtained thermal-IR images of W49A at wavelengths of 8.8, 9.7,
11.6, and 18.5 $\mu$m using Michelle on the Gemini North telescope.
Table~1 lists observation dates and other details.  Michelle is the
facility mid-IR imager and spectrograph on Gemini North, with a
320$\times$240 pixel Si:As IBC array, a pixel scale on the 8~m Gemini
North telescope of 0$\farcs$099, and a resulting field of view of
31$\farcs$7$\times$23$\farcs$8.  The observations were taken with a
$\sim$15\arcsec\ north-south chop throw.  W49A is a complex, bright
extended source, so some fainter emitting sources contaminated the
reference sky positions, leading to negative image residuals in the
chop-nod data.  These artifacts were corrected as well as possible
using the same technique that we used previously on a large mosaic of
the bright inner regions of the Orion Nebula (see Smith et al.\ 2004,
2005).  We subdivided the pixels without interpolation in each of the
various images and then shifted and co-added them, to produce a final
mosaic image at each of the four wavelengths we observed.  The
measured FWHM in our final registered and co-added 8--12 $\mu$m images
was 0$\farcs$3--0$\farcs$35, and about 50\% larger at 18.5 $\mu$m,
roughly consistent with the expected diffraction limit.

Figure~\ref{fig:one} shows the resulting co-added Michelle images.
The observations were performed in service mode and targeted the water
maser source W49A/G, but the $\sim$30\arcsec$\times$20\arcsec\
field-of-view included a few other sources in the brightest central
and western parts of the ring of UCHII regions as well (Welch et al.\
1987).  The observing conditions were not optimal; while the image
quality was good, the sky was non-photometric, and suitable standard
star observations were not obtained.  The mid-IR images presented here
could therefore not be absolutely flux calibrated and our discussion
must be limited mainly to the observed morphology.  Our previous
ground-based observations, however, have already measured the mid-IR
fluxes of various sources in our field (Smith et al.\ 2000), and we
incorporate these and other available photometric data in our models
described in \S 4.4.  In principle, the bandpasses of the 8.8 and 11.6
$\mu$m filters allow for partial contribution of PAH (polycyclic
aromatic hydrocarbon) features.  However, a low-resolution 8--13
$\mu$m spectrum of source G published by Gillet et al.\ (1975) shows
only smooth continuum emission with deep 9.7~$\mu$m silicate
absorption, so we suspect that our images are not strongly influenced
by extended PAH emission.  They may, however, contain some extended
silicate emission (see below).


\begin{table}\begin{minipage}{3.4in}
\caption{Gemini North/Michelle Observations of W49A}\scriptsize
\begin{tabular}{@{}lrcl}\hline\hline
Obs.\ Date &$\lambda$($\mu$m) &$\Delta\lambda$($\mu$m)  &Exp.\ Time (s) \\ \hline
%
2004 Aug 11 &11.6 &1.1  &6$\times$155  \\
2004 Aug 11 &18.5 &1.6  &2$\times$82   \\
2004 Sep 24 &11.6 &1.1  &6$\times$155  \\
2004 Sep 24 &18.5 &1.6  &6$\times$82   \\
2004 Sep 30 &8.8  &0.9  &3$\times$207  \\
2004 Oct 5  &8.8  &0.9  &6$\times$259  \\
2004 Oct 5  &9.7  &1.0  &6$\times$221  \\
2004 Oct 5  &11.6 &1.1  &1$\times$52   \\
\hline
\end{tabular}
\end{minipage}
\end{table}

\subsection{Multiwavelength Registration}

Aligning the mid-IR images to one another at the four wavelengths was
relatively simple; registering to a common compact source such as
source F provided satisfactory results (a color image showed no
perceptable registration problems between IR wavelengths).  However,
we also wish to compare our new mid-IR images to high-resolution radio
continuum data (Fig.\ 2$a$).  The absolute positional accuracy of our
images ($\sim$1\arcsec) is not sufficient to provide a meaningful
comparison between the high-resolution IR and radio data, and
registering to a common source may be problematic since we are dealing
with thermal dust emission in the mid-IR and ionized gas in the radio.
We show two alternative solutions to the relative IR/radio
registration in Figure 2.

1. In Figure 2$b$, the radio and IR data are aligned by matching the
centroid position of source F, which appears point-like in the radio
at this resolution.  Source F is the only one detected in the
field-of-view of our IR images whose photospheric emission is detected
in the near-IR $K$ band (Conti \& Blum 2002; Alves \& Homeier 2003).
Under this assumption, the precision to which we can align the images
would be $\sim$0$\farcs$01 (roughly 10\% of a pixel) based on the
centroiding precision of the point source, although we suspect that
this assumption is incorrect because source F shows complex extended
structure in the mid-IR.

2. The second registration option in Figure 2$c$ and 2$d$, which we
favor, uses a cross correlation of sources I and J for registering the
radio and IR images.  The precision of the cross-correlation between
the radio and IR data is $\sim$0$\farcs$05 in both R.A.\ and DEC.  The
cross correlation was performed between the radio and both 11.6 and
18.5 $\mu$m images for both sources I and J, and the cross correlation
function was well-behaved (i.e. single-peaked Gaussian).  At the
angular scales sampled here we do not spatially resolve dust
temperature gradients within the thin walls of cometary or shell-like
UCHII regions such as I and J, so we expect the mid-IR morphology to
roughly trace that in the radio continuum, since the dust emitting at
10--20 $\mu$m is heated predominantly by trapped Ly$\alpha$ photons
(e.g., Hoare et al.\ 2007; Smith \& Brooks 2007). Indeed, most of the
mid-IR emission detected from UCHII regions in W49 shows very good
spatial and morphological agreement with the radio continuum (Smith et
al.\ 2000), as is the case for the cometary UCHII region G29.96-0.02
(De Buizer et al.\ 2002).

{\it Which of these two is correct?}  These two options differ in
alignment by about 0$\farcs$6, mainly in the east-west direction, and
bear upon the location of the IR peak of source G compared to the
location of the water maser outflow's origin, as discussed later.  In
option 1, the compact source F is nicely aligned, but the
mis-alignment of the extended radio and IR sources I and J looks
rather strange.  Moreover, the IR centroids of I and J would,
coincidentally, need to be offset from their radio conterparts by the
same amount and in the same direction. Option 1 would locate the main
IR source G:IRS1 overlapping with G1 (the western peak of source G),
which seems problematic since the well-defined shell-like radio
morphology of G1 (see below) does not correspond well with the
centrally-peaked and possibly bipolar morphology of G:IRS1.  Option 2
seems better, as long as it is reasonable to assume that source F in
the radio does not exactly match the position of source F in the
mid-IR.  This may be true, because despite its point-like radio
appearance and stellar detection in the near-IR, source F is not a
pure point source in our new mid-IR Gemini images anyway.  It shows
some faint extended structure around a bright core.  

As a final check, we compared the radio position of source F to the
absolute position of the mid-IR source in data obtained with the
Infrared Array Camera (IRAC) in the {\it Spitzer Space Telescope}
during the GLIMPSE survey (Benjamin et al.\ 2003).  The absolute
coordinates in the GLIMPSE data are determined with reference to
2MASS, and are accurate to 0$\farcs$3, so while the {\it Spitzer}
images have lower angular resolution than our new Gemini data, they
are sufficient top solve the registration ambiguity between options 1
and 2 above.  In the IRAC band 3 (5.8 $\mu$m) filter, the position of
source F is $\alpha_{2000}$ = 19:10:13.39, $\delta_{2000}$ =
+9:06:21.56, whereas DePree et al.\ (2000) give the radio continuum
position of source F as $\alpha_{2000}$ = 19:10:13.345,
$\delta_{2000}$ = +9:06:21.47.  From this comparison, the mid-IR
counterpart of source F is offset by 0$\farcs$6 east and 0$\farcs$09
north of the radio position. (IRAC bands 1 and 2 yield similar
offsets, whereas the IRAC band 4 data are affected by severe
over-exposure of source G and have artifacts across the image.)  This
offset is consistent with our option 2 registration discussed above,
and allows us to rule out option 1.  Interestingly, the near-IR
$K$-band point source and an X-ray source associated with F also
appear to be slightly offset to the east of the radio continuum source
by roughly 1\arcsec\ (Tsujimoto et al.\ 2006), nearly coincident with
the mid-IR source in option 2.  Thus, we adopt option 2 below in the
discussion of Source G.

\section{RESULTS FOR SOURCES OTHER THAN  G}

\subsection{The Missing IR Sources in the Ring and the Environment of
  Source G}

Smith et al.\ (2000) presented the first mid-IR survey of warm
dust emission from the UCHII regions in W49A.  Nearly all of the
known radio continuum sources thought to be UCHII regions were
clearly detected in the mid-IR images, and when resolved, the radio and
IR morphologies generally matched.

This general agreement between IR and radio emission accentuates the
fact that the few radio sources that were {\it not} detected in the IR
are all clustered together within $\sim$5\arcsec\ of one another at
the western end of the ring.  These missing IR sources were W49A/A, B,
C, D, and E, plus their associated subcomponents.  In terms of
extinction, there appeared to be a sharp vertical dividing line at
$\alpha$(2000)=19$^h$10$^m$13$\fs$25 (see Fig.~\ref{fig:two}), such
that no sources in the ring were detected west of this line in the IR.
Smith et al.\ (2000) estimated a lower limit of roughly $N_{H_2} >
10^{23}$ cm$^{-2}$ for the column density required to fully extinguish
these sources at 20$\mu$m.  In fact, several observations in tracers
of dense molcular gas (like NH$_3$, SO$_2$, C$^{34}$S, etc.) found
higher concentrations toward the western part of the ring (Jackson \&
Kraemer 1994; Serabyn et al.\ 1993, Dickel \& Goss 1990), and Dickel
\& Goss (1990) estimated the line of sight molecular column density
toward source A to be $N_{H_2} = 2.5 \times 10^{24}$ cm$^{-2}$.  This
is more than sufficient to completely extinguish these sources in the
mid-IR at the sensitivity limits of Smith et al.\ (2000).

Our new observations in Figure~\ref{fig:one} confirm the non-detection
of sources A--E, including additional wavelengths shorter than
12~$\mu$m that were not observed in our previous paper.  Although our
new Gemini images at 10--20 $\mu$m are not absolutely flux calibrated,
we can use the photometry in Smith et al.\ (2000) to bootstrap from
the observed counts for sources that are detected in our images (like
sources F, I, and J).  The diffuse background level in our new images
is about the same as in Smith et al.\ (2000) at 12--20 $\mu$m, but we
are about two times more sensitive to compact unresolved sources
because of the higher angular resolution of the Gemini images.  Thus,
the column density required to extinguish these sources is even higher
than estimated in our previous paper, but is still plausible given the
observed molecular column density.  Of course, the local column
density to any of the sources may in fact be higher than estimated
from molecular observations if they are clumped on size scales smaller
than the beam.

Sources A--E also tend to be among the most compact UCHII regions in
W49.  Sources G, I, and J in the middle of the ring are more extended,
while Source L at the far eastern side of the ring is very diffuse.
Combining this clue with the much higher molecular column density and
the non-detection in the thermal-IR, Smith et al.\ (2000) proposed
that these missing IR sources at the western end of the ring
constitute the youngest and most embedded UCHII regions in W49, and
that there seems to be a west-to-east age gradient in W49.
Subsequently, Alves \& Homeier (2003) reported the discovery of a
massive young cluster of O-type stars just 3~pc east of the ring,
supporting this overall picture of an east/west age gradient.  Alves
\& Homeier note that this may not necessarily imply triggering of star
formation in the ring by the existing O-star cluster, since there
appear to be other seeds of star formation elsewhere in W49, such as
W49 South and sources S, R, and Q (see Smith et al.\ 2000; De Pree et
al.\ 1997).  Moving from east to west, one sees the massive O-star
cluster, followed by diffuse emission in a strong ionization front
seen in near-IR mages (including Br$\gamma$) and deep radio continuum
images, followed by the Welch ring.  This line of reasoning suggests
that the western end of the ring of UCHII regions is the densest,
youngest, and most active part of the W49 cloud core. 

Its location right at the boundary of a dense cloud core bears upon
source G's status as the most powerful water maser outflow source in
the Milky Way.  Its location hints that the observable onset of this
outflow phase may be related to the density structure of the
surrounding environment as well as the age of the protostellar outflow
source itself.  In this context, it is intriguing that the ionized
region and outflow around source G is more extended to the east than
to the west, as discussed further below.

\begin{figure}\begin{center}
\epsfig{file=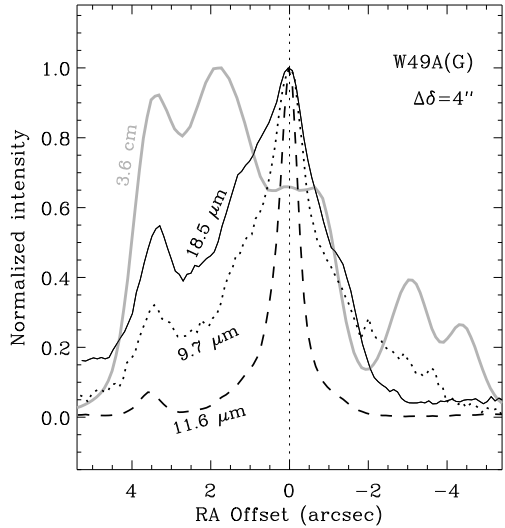,width=3in}
\end{center}
\caption{Intensity tracings at IR wavelengths and the radio continuum
  through source G using a 4\arcsec-wide sample in declination.  The
  extended emission around source G is brighter toward the east of the
  IR peak.}\label{fig:three}
\end{figure}

\subsection{Sources F,  I,  and J}

In our previous ground-based IR imaging of W49 with a 3~m telescope,
source F was unresolved, whereas I and J were possibly extended with
unclear morphology (Smith et al.\ 2000).  Each is powered by the
equivalent ionizing flux of an O6.5--O7.5 V star (De Pree et al.\
1997).  With the better diffraction limit in our new Gemini mid-IR
images (Fig.~\ref{fig:one}), all three show clear evidence for
extended structure.  Sources I and J show no compact IR source,
appearing instead as cometary or shell-like UCHII regions.  Their
thermal-IR morphologies are well-matched by the diffuse structure seen
in the radio continuum (see Figs.\ 2$c$ and 2$d$).

Source F is more complicated.  It appears slightly resolved at
8.8~$\mu$m, with a FWHM of 0$\farcs$5.  This is more extended than the
angular resolution of the image (source G has a FWHM of roughly
0$\farcs$3 in the same 8.8~$\mu$m frame).  Moving to longer
wavelengths, Source F develops an unresolved point-like core with an
extended halo.  This halo is asymmetric, being more extended toward
the east, and the asymmetry is most pronounced at 18.5 $\mu$m in
Figure~1$d$.  Source F is very compact in the radio continuum (it is
the most compact 3.6~cm source in the ring observed by De Pree et al.\
1997); it appears as a point source in the 3.6~cm image in Figure
2$a$. In our favored alignment of the IR and radio images (Figs.\ 2$c$
and 2$d$), the IR source is offset from the peak of the 3.6~cm
continuum emission by $\sim$0$\farcs$6 at P.A.=74$\pm$3\arcdeg.  This
is the same direction toward which the diffuse halo of F is most
extended at 18.5~$\mu$m.  This slight offset is unusual but not
unprecedented; several compact regions studied by De Buizer et al.\
(2005) show similar small offsets between compact thermal-IR sources
and the centimeter-wavelength radio continuum.  The difference between
IR and radio positions of source F is of interest because it is the
only source we have detected in the ring for which stellar
photospheric emission has been detected in the near-IR (Conti \& Blum
2002; Alves \& Homeier 2003).  Blum (2005) presented a near-IR
spectrum of source F, showing strong Pa$\alpha$ and Br$\gamma$
emission and a very red continuum.

The mid-IR/radio offset we observe here and the extension of diffuse
IR emission toward the east implies very high local extinction.  Light
escaping eastward out of the source fits the overal trend of higher
extinction as one moves westward within the ring of UCHII regions.
Consequently, the near-IR source associated with F may not be the same
star that ionizes the gas detected in the radio continuum.  Indeed,
the position of the near-IR source is also offset to the east of the
radio continuum source. 
Interestingly, a hard X-ray source was recently detected in the
vicinity of source F (Tsujimoto et al.\ 2006).  It is located even
further west than the radio source, adding to our suspicion that F may
harbor multiple protostellar sources or complex substructure.

The UCHII regions K, L, and M are outside our field of view.  We did
not detect source H because its surface brightness is too low in the
thermal-IR (Smith et al.\ 2000).

\begin{figure}\begin{center}
\epsfig{file=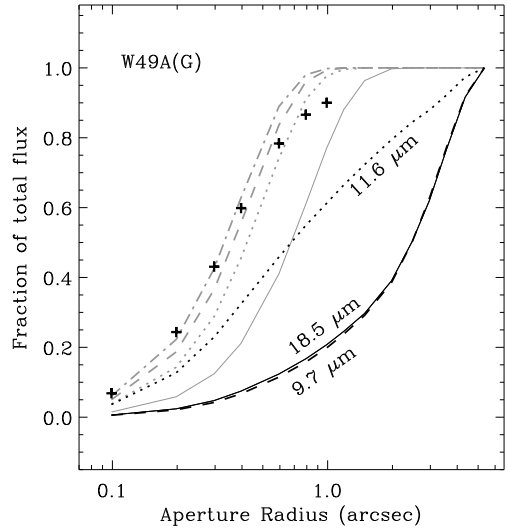,width=3in}
\end{center}
\caption{Encircled fraction of the total flux for source G in the
  Michelle images.  The profiles for 18.5 $\mu$m (solid) and 9.7
  $\mu$m (dashed) are identical, despite the differences in angular
  resolution.  The 11.6 $\mu$m profile (dotted) shows more centrally
  concentrated flux, but is still an extended source.  The profile for
  8.8 $\mu$m (plus signs) is less definite because of low signal to
  noise in the outer wings of the PSF; it is almost consistent with an
  unresolved PSF where only about 10\% of the emission is extended.
  Profiles for the encircled flux of a diffraction-limited PSF at each
  wavelength are shown in gray for comparison.}\label{fig:four}
\end{figure}

\section{DETAILED MORPHOLOGY OF THE OUTFLOW SOURCE G}

\subsection{Apparent Structure in IR Images}

Source G shows intriguing extended structure in Figure~\ref{fig:one}.
Previous mid-IR observations indicated that it was extended out to a
radius of $\sim$5\arcsec, but had insufficient resolution to uncover
its small-scale structure (Smith et al.\ 2000).

The bright central peak, which we denote source G:IRS1, appears to be
at the center of an X-shaped distribution of extended IR emission
reaching out to $\sim$2\arcsec\ from the central peak in our 9.7,
11.6, and 18.5~$\mu$m images (the extended emission was not detected
at 8.8~$\mu$m because of lower sensitivity).  The central peak itself
has a slight cometary shape on a scale of $\sim$0$\farcs$5, opening
toward the east and connecting to part of the larger X shape.  This
morphology is characteristic of a flared disk geometry that is viewed
from an intermediate angle, and is seen in the central sources in the
dust emission models we present later in \S 4.4.  Beyond the central
peak and X-shaped nebula, source G shows complex, multiple-peaked
structure in a ``halo'' that extends out to $\sim$5\arcsec\ from the
center, and seems to have a larger extent at 9.7 and 18.5~$\mu$m than
it does at 11.6~$\mu$m.  In particular, there is a second resolved
source $\sim$3$\farcs$5 east of G:IRS1, nearly coincident with the
radio continuum source G5 (see Fig.~\ref{fig:one}).  It appears to be
a point source at 8.8--18.5 $\mu$m, although in the radio continuum G5
seems to be part of a limb brightened cavity wall.  This IR source may
represent a second IR protostar forming near G:IRS1, or it may simply
be a condensation in the cavity wall.

Following the systematic trend of higher densities toward the west in
the ring of W49, the IR halo of source G is more extended toward the
east and ends more abruptly on its west side (Fig.~\ref{fig:three}).
The fact that the radio continuum follows this trend as well
(Fig.~\ref{fig:three}) means that it is not a mere extinction gradient
along our line of sight, but rather, a true physical density gradient
in the surroundings, making it easier for photons to escape toward the
east, or for outflows to carve cavities in that direction.

\begin{figure}\begin{center}
\epsfig{file=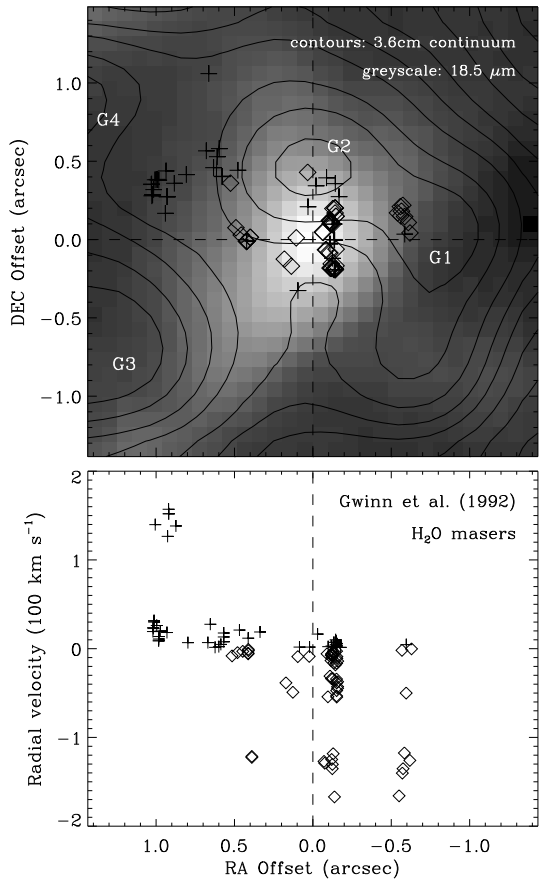,width=2.8in}
\end{center}
\caption{The top panel shows 3.6~cm contours and the H$_2$O maser
  positions from Gwinn et al.\ (1992) plotted over the 18.5 $\mu$m
  Gemini/Michelle image in grayscale.  The bottom panel shows the
  distribution of H$_2$O masers (Gwinn et al.\ 1992).}
\label{fig:masersZOOM}
\end{figure}

\begin{figure}\begin{center}
\epsfig{file=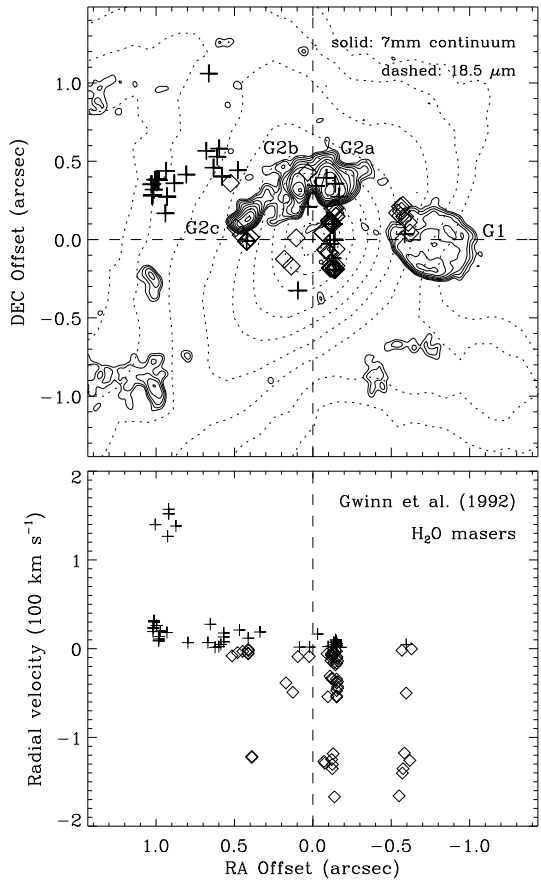,width=2.8in}
\end{center}
\caption{Same as Figure 5, except that the 18.5 $\mu$m image is shown
  in dashed contours, and solid contours show the high resolution 7~mm
  continuum image from De Pree et al.\ (2000) instead of the 3.6~cm
  data.}
\label{fig:masers7mm}
\end{figure}

Figure~\ref{fig:four} confirms the general impression from raw images
that the 9.7 and 18.5~$\mu$m images are more spatially extended than
the 11.6~$\mu$m emission.  The encircled flux profiles of the 9.7 and
18.5~$\mu$m filters are nearly identical, while the 11.6 $\mu$m
profile has a different shape.  The 9.7 and 18.5 $\mu$m~filters sample
silicate emission or absorption features, while the 11.6~$\mu$m filter
is dominated by warm dust continuum emission in W49 (Gillet et al.\
1975).  Thus, the similar radial profiles of the 9.7 and 18.5~$\mu$m
filters in Figure~\ref{fig:four} suggest that both filters sample
extended silicate emission, and may also be affected by heavier
silicate absorption of the continuum toward the central source.  The
silicate emission extends over a region more than 0.5 pc across, while
the continuum source at 11.6~$\mu$m is more centrally concentrated.
This may suggest UV excitation of the silicate emission by multiple
stellar sources in the vicinity of source G.  Spatially-resolved
mid-IR spectroscopy of this extended emission would be worthwhile to
confirm this conjecture.

%





\subsection{Identification of the H$_2$O Maser Outflow Source and the
  Hot Molecular Core}

Figure~\ref{fig:masersZOOM} shows the environment immediately
surrounding source G:IRS1.  This figure includes 3.6 cm radio
continuum contours using the same alignment as in Figures 2$c$ and
2$d$, for which the registration accuracy is roughly 0$\farcs$05, as
noted earlier.  The registration of the water masers compared to the
radio continuum was adopted from the study of De~Pree et al.\ (2000),
with a quoted accuracy of $\sim$0$\farcs$05.  Therefore, the accuracy
of the registration between the IR image and the water masers is
roughly 0$\farcs$07 with the positional uncertainties added in
quadrature.  Radial velocities and E/W positional offsets of the water
masers from Gwinn et al.\ (1992) are plotted in
Figure~\ref{fig:masersZOOM} as well (bottom panel).  The reference
position from which offsets in R.A.\ and DEC are measured here
corresponds to the center of the maser outflow in the favored best-fit
model of Gwinn et al.\ (1992); specifically, it is the presumed origin
point for their ``solution 4'' listed in their Table~5.

It is clear that the expected launching source of the water maser
outflow is coincident with G:IRS1 to within the positional error of
our study, whereas G:IRS1 and the water maser source are both
significantly offset from the nearest 3.6~cm radio continuum peak G2
at this resolution.  G:IRS1 is located about 0$\farcs$4 due south of
the 3.6~cm centroid of G2 (Figure 5$a$).\footnote{If instead we had
  adopted our alternative choice for the registration of IR and radio
  images using source F for the registration (see Fig.\ 2$b$), source
  G:IRS1 would be located approximately coincident with radio
  continuum source G1.  This seems unlikely because the clear
  shell-like radio morphology of source G1 (see
  Fig.~\ref{fig:masers7mm} and De Pree et al.\ 2000) has no
  correspondence with the observed IR morphology.}  We found no
suitable image registration that would align G:IRS1 with radio source
G2.

Figure~\ref{fig:masers7mm} is the same as Figure~\ref{fig:masersZOOM},
except that it compares the 18.5~$\mu$m Gemini image (dashed contours)
to a higher resolution 7~mm radio continuum image from De~Pree et al.\
(2000).  Again we see that the IR peak is offset
0$\farcs$35--0$\farcs$4 south from the strongest emission associated
with G2a/b.  Source G1 appears to be a relatively isolated, unrelated
shell-like UCHII region associated with no mid-IR source.

\begin{figure}\begin{center}
\epsfig{file=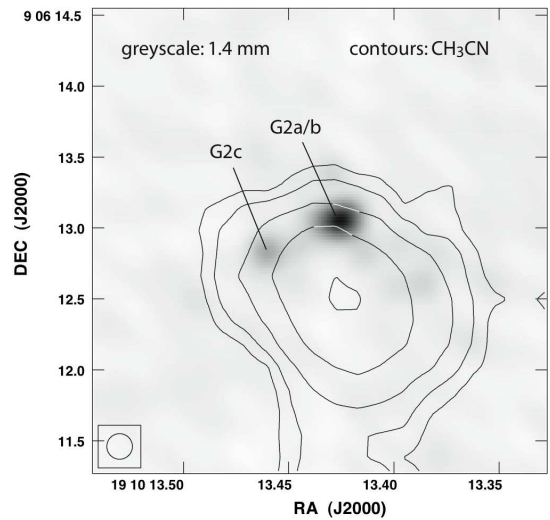,width=2.8in}
\end{center}
\caption{A similar field of view as in Figure~\ref{fig:masers7mm},
  showing the 1.4 mm continuum emission (greyscale) from sources G2a/b
  and G2c, and CH$_3$CN line emission (contours) from the hot core.
  These data were presented originally by Wilner et al.\ (2001), and
  were obtained with the BIMA (Berkeley Illinois Maryland Association)
  array. The continuum and line emission are from the same dataset, so
  there is no error in the registration, demonstrating that the hot
  core is reliably centered $\sim$0$\farcs$5 south of G2a/b.  The
  molecular hot core emission is coincident with our new source G:IRS1
  (compare with Fig.\ 6).}
\label{fig:hotcore}
\end{figure}

\begin{figure}\begin{center}
\epsfig{file=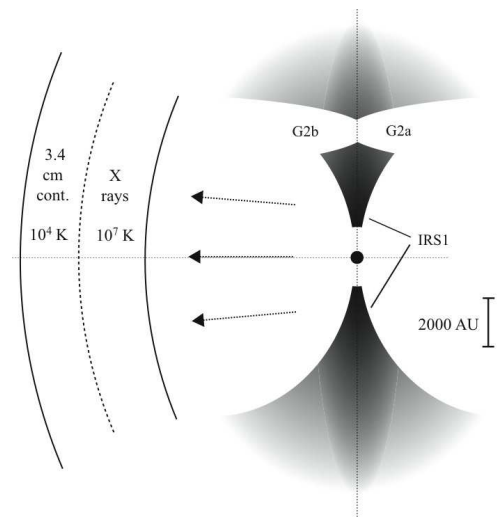,width=2.8in}
\end{center}
\caption{A hypothetical picture of the relationship between the hot
  core G:IRS1 and the UCHII region G2a/b (see text).}
\label{fig:sketch}
\end{figure}

If our spatial alignment is correct, then another interesting
coincidence is that the location of G:IRS1 matches the position of a
hot molecular core seen in CH$_3$CN emission (Wilner et al.\ 2001;
their source``b'' located 0$\farcs$1 to 1\arcsec\ south of G2).
Figure~\ref{fig:hotcore} shows the emission from this hot core
(contours of CH$_3$CN emission) compared to the the 1.4 mm continuum
from sources G2a/b and G2c in the data presented by Wilner et al.\
(2001).  These line and continuum measurements are from the same
dataset, so there is no registration error in the relative positions.
Figure~\ref{fig:hotcore} shows that the molecular hot core source is
clearly offset south of G2a/b, making it coincident in location and
size with our new IR source G:IRS1.  By analogy with the hot core and
H$_2$O maser outflow in Orion, this strengthens the case that G:IRS1
(and not G2a/b) is in fact the source of the water maser outflow in
W49.  Combined with the lack of a compact radio continuum source and
the presence of infall indicated by inverse P Cygni profiles in CS
(Williams et al.\ 2004), G:IRS1 appears to be an excellent candidate
for an accreting massive protostar in a hot molecular core accompanied
by a bipolar outflow.  This differs from the case of the cometary
UCHII region G29.96--0.02, for example, where the mid-IR source is
offset from the hot molecular core and coincident with the radio
continuum instead (De Buizer et al.\ 2002).

\begin{figure*}\begin{center}
\epsfig{file=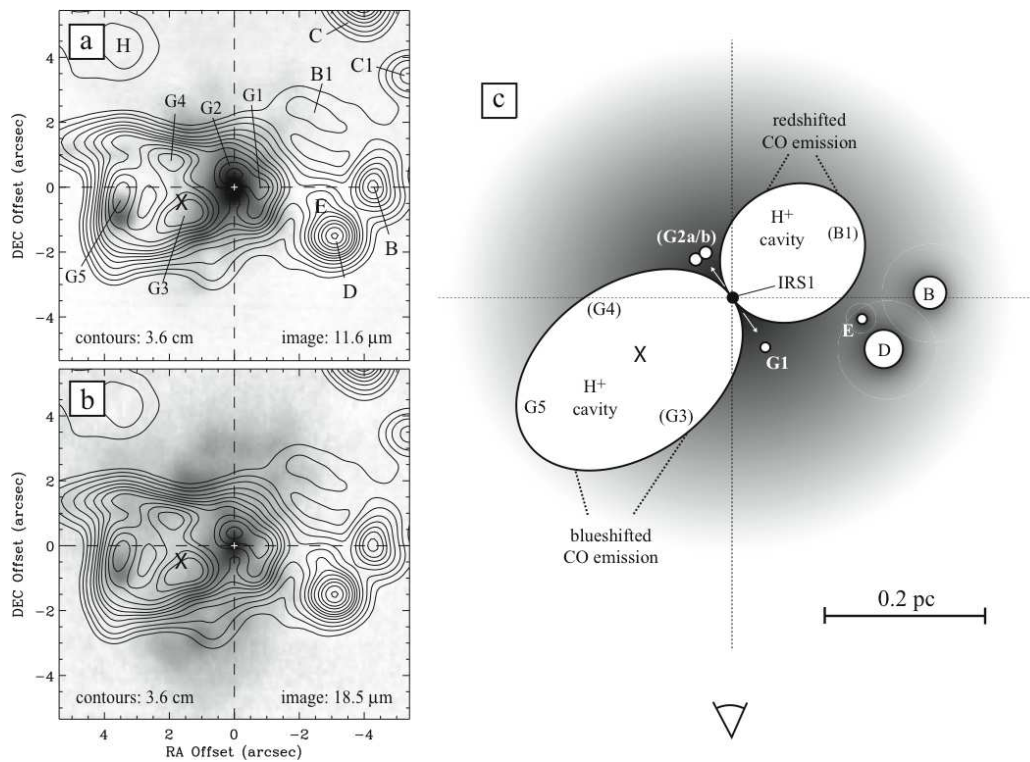,width=5.4in}
\end{center}
\caption{Contours of 3.6~cm radio continuum superposed over the
  Gemini/Michelle images of source G at 11.6~$\mu$m (a) and
  18.5~$\mu$m (b). (c) A cartoon of one possible geometric model for
  source G and its environment (see text), viewed from the north (an
  Earth-based observer is at the bottom of the drawing).  Features
  which are well above or below the plane of the drawing are in
  parentheses.  The walls of the large cavities give rise to the
  blueshifted and redshifted CO outflow to the east and west,
  respectively (Scoville et al.\ 1986), while the white arrows in the
  equatorial plane denote the approximate zones where the H$_2$O
  masers are found.  The ``X'' marks the spot associated with the
  approximate peak and centroid of hard X-ray emission detected
  recently by Tsujimoto et al.\ (2006).}
\label{fig:outflow}
\end{figure*}

The water masers are spread across a 2\arcsec\ range on either side of
G:IRS1, and are elongated in a primarily east/west orientation
following the presumed direction of the maser outflow (Gwinn et al.\
1992).  The tightest cluster of maser spots is found just a bit more
than 0$\farcs$1 west of G:IRS1.  This maser cluster has a clearly
linear arrangement, and is elongated along a north/south axis
perpendicular to the larger outflow.  Essentially all the masers in
this cluster are blueshifted (there is one very low velocity
redshifted spot), and the feature is persistent over decades (Walker
et al.\ 1982; Gwinn et al.\ 1992; De Pree et al.\ 2000).  If the water
masers form at the edge of the cavity walls, as suggested by Mac Low
\& Elitzur (1992) and Mac Low et al.\ (1994), then this linear
arrangement of predominantly blueshifted masers might represent a
position where our line of sight skims a tangent point in the wall of
the blueshifted outflow cavity or flared edges of a disk (see Figures
\ref{fig:sketch} and \ref{fig:outflow}, and discussion below).  Such
linear arrangements are not unusual.  In several cases where water
masers are closely associated with UCHII regions detected in the
mid-IR, the masers show a quasi-linear distribution on the sky (De
Buizer et al.\ 2005).  De Buizer et al.\ (2005) conclude that in most
cases the linear distribution traces the outflows and not the disks.

Although G:IRS1 and G2a/b are not coincident and are not powered by
the same central (proto)star, they may have a ``symbiotic''
relationship.  Sources G2a and b are closely connected, being bridged
by faint emission, and may be part of the same bipolar structure.  The
pinched waist that divides them is projected along the same
north/south line that passes through G:IRS1 and is perpendicular to
the large-scale outflow axis.  Given the apparent bipolar morphology
of source G2a/b, which is oriented along the same outflow axis, it is
concievable that both sources are aligned and that they may have
formed from the same flattended rotating cloud core that fragmented as
it collapsed, as depicted in Figure~\ref{fig:sketch}.  Kratter \&
Matzner (2006) argue that massive protostellar disks with the
accretion rate we infer for G:IRS1 (see below) will be unstable to
fragmentation, although on smaller size scales of $\sim$150 AU, while
numerical simulations by Krumholz et al.\ (2009) show multiple stars
forming from a common rotating disk/envelope with characteristic
separations of $\sim$1000 AU.

In any case, if they share a real physical proximity, the source that
ionizes G2a/b may also help to ionize source G and its outflow on a
larger scale; G2a/b requires an ionizing flux that is the equivalent
of an O5.5 V star (De~Pree et al.\ 2000).  With such a luminous star,
it is puzzling that we detect no mid-IR emission from G2a/b;
additional extinction along the line of sight or cooler dust
temperatures located farther from the star may play a role in this
mystery. This picture is of course still very speculative, but it is
compelling enough to suggest that the close relationship between
G:IRS1 and G2a/b deserves continued study as it may provide direct
insight to the formation of massive binary systems.  In this context,
Source G in W49 may therefore be an excellent target to observe with
{\it ALMA}.

\begin{table*}\begin{center}\begin{minipage}{3.4in}
\caption{Models for the IR emission from G:IRS1}\scriptsize
\begin{tabular}{@{}lcccccl}\hline\hline
  Model &$M_*$  &$L_*$  &$\dot{M}_{acc}$ &$i$  &$\theta$ &comment \\
  &($M_{\odot}$) &(10$^5$ $L_{\odot}$) &(10$^{-4}M_{\odot}$ yr$^{-1}$) &(deg) &(deg) & \\
  \hline

A &25--35 &1--2 &1--10 &85     &11-18 &fits silicate abs., not SED or image  \\
B &25--35 &2--3 &5     &30--90 &2--3  &fits SED, not silicate abs.\ or image  \\
C &45     &3    &10    &60     &30    &fits silicate abs., SED, and image  \\

\hline
\end{tabular}
Note: $i$ is the inclination angle at which we view the system, and
$\theta$ is the outflow opening angle.
\end{minipage}\end{center}
\end{table*}

\subsection{The Outflow Geometry of Source G: A Bipolar Cavity}

Figures~\ref{fig:outflow}$a$ and \ref{fig:outflow}$b$ compare the
3.6~cm radio continuum morphology to that seen in our 11.6 and
18.5~$\mu$m Gemini images.  Clearly, the large-scale radio continuum
is more elongated east/west, while the diffuse IR emission at
18.5~$\mu$m seems to be elongated in the opposite direction.  The
wings of the X-shaped IR nebulosity that emerge to the NE and SE from
G:IRS1 seem to outline the brightest diffuse radio continuum emission
in the eastern part of G (sources G3 and G4).  It seems likely that
the mid-IR emission comes from warm dust at the limb-brightened
boundary of an ionized cavity; our model images discussed in the next
section show that this interpretation is plausible.

The large cavity is likely to be a stellar wind-blown bubble (e.g.,
Weaver et al.\ 1977), since it is partly filled with X-ray emission.
The approximate center of the extended X-ray sources detected by
Tsujimoto et al.\ (2006), which has a similar extent to the radio
emission, is marked by the ``X'' in Figures \ref{fig:outflow}$a$ and
\ref{fig:outflow}$b$.  In that case, the hard X-rays are likely to be
the hot stellar wind or jet decelerated in the reverse shock (see
Fig.~\ref{fig:sketch}, for example).  A similar geometry may apply to
the western lobe of source G, where source B1 and other low-level
diffuse radio emission may define a similar -- but smaller -- ionized
cavity on the opposite side, which is bounded close to the source by
the fainter IR extensions to the SW and NW from G:IRS1.  The western
side of the putative bipolar outflow is more complicated, because
there are other UCHII regions (sources B, D, and E) projected along
the same line of sight to the redshifted lobe, and because the diffuse
radio emission from the redshifted lobe is fainter.  We are therefore
less confident about the receding lobe's geometry, although its
smaller size makes intuitive physical sense if its environment is
denser.

A basic geometric model is sketched in Figure \ref{fig:outflow}$c$.
Tilting the eastern (left in Fig.~\ref{fig:outflow}$c$) cavity toward
the observer is favored by several observational properties: 1) the
eastern lobe is brighter, 2) the UCHII regions to the west are all
obscured in the mid-IR (Smith et al.\ 2000), 3) the morphology of
G:IRS1 is consistent with the eastern polar axis being tilted toward
us, and 4) this is the orientation of the $\sim$25 km s$^{-1}$ CO
outflow observed on similar size scales by Scoville et al.\ (1986),
where the blueshifted CO emission is toward the east and the
redshifted CO emission is offset to the west.  The CO outflow probably
traces a dense sheath or cocoon surrounding the ionized cavity.

This orientation for the CO outflow and ionized cavities is orthogonal
to the water maser outflow (the small white arrows in
Fig.~\ref{fig:outflow}$c$).  One solution to this discrepancy, as
noted above, may be that the water masers trace dense material in an
equatorial torus or envelope, or may originate at the interface
between the cavity and disk envelope close to the source (Mac Low et
al.\ 1994).  A similar orientation is seen in the OMC-1 core (e.g.,
Plambeck et al.\ 1982; Greenhill et al.\ 1998), and also for S140 IRS1
(Hoare 2006).

From Figure \ref{fig:outflow}$c$ it is easy to picture how the linear
water maser feature located 0$\farcs$1 west of G:IRS1 could arise near
the tangent point of the blueshifted outflow cavity.  The overall
outflow geometry for source G that is pictured in
Figure~\ref{fig:outflow}$c$ is nearly identical to that envisioned
earlier by Dickel \& Goss (1990), although their picture was based
primarily on the different column densities observed toward various
lines of sight.  In fact, Dickel \& Goss argued that because of an
abrupt change in column density, the edge of the blueshifted part of
the outflow must reside between sources G and D, as we have pictured
here and have associated with the linear water maser feature.

IRS1 is not concident with the ionized UCHII regions G2a, G2b, or G1
(Fig.~\ref{fig:sketch}), showing that there are several centers of
massive star formation activity even on size scales of $\sim$0.1
pc. Thus, in priciple, we do not necessarily need the water maser
outflow to be aligned with the larger cavity because it could simply
be due to a different outflow with a different axis orientation.  For
example, in OMC-1, the large-scale molecular outflow and cavity traced
by mid-IR emission and the shocked H$_2$ emission from the system of
``fingers'' defines one axis that runs SE to NW (Allen \& Burton 1993;
Kaifu et al.\ 2000; Kwan \& Scoville 1976; Gezari et al.\ 1998;
Shuping et al.\ 2003; Smith et al.\ 2005), while source IRc2 seems to
drive a smaller collimated outflow nearly perpendicular to the large
scale flow (Greenhill et al.\ 1998, 2003; Bally et al.\ 2005).  The
OMC-1 South core has several jets and molecular outflows in various
directions that all originate in a $\sim$0.01 pc region with multiple
mid-IR sources (Smith et al.\ 2004; Zapata et al.\ 2004).

\subsection{Radiative Transfer Modeling}

\begin{figure}\begin{center}
\epsfig{file=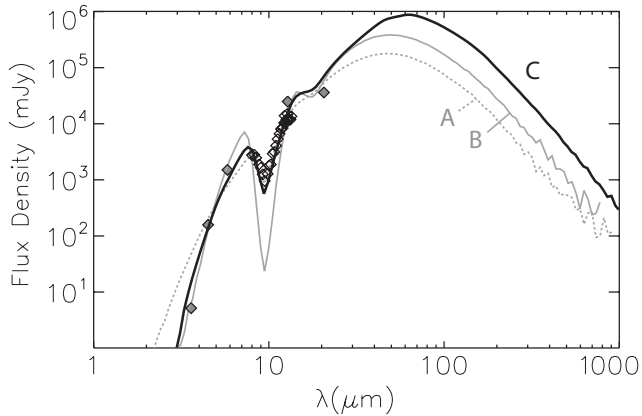,width=3.3in}
\end{center}
\caption{The mid-IR spectral energy distribution of source G in W49
  along with protostellar envelope dust emission models.  The unfilled
  diamonds are from Gillet et al.\ (1975), and were obtained with a
  large 22\arcsec\ diameter aperture.  The solid diamonds are imaging
  photometry from Smith et al.\ (2000) and {\it Spitzer}/IRAC
  photometry from the GLIMPSE project.  Best-fitting model SEDs are
  shown.  Models A and B (gray) correspond to fits to the SED alone,
  whereas model C (black) takes into account the opening angle and
  morphology in images (see text, \S 4.4). Model C corresponds to a
  star of mass 45 $M_{\odot}$ and luminosity 3$\times$10$^5$
  $L_{\odot}$ with an envelope accretion rate of 10$^{-3}$ $M_{\odot}$
  yr$^{-1}$ and an inner cavity of radius 1700 AU. }
\label{fig:SED}
\end{figure}

In order to verify that the morphology and orientation of the bipolar
cavity are generally correct as sketched in
Figure~\ref{fig:outflow}$c$, and to constrain physical properties of
the illuminating source, we conducted simulations of the dust emission
to compare with the observed mid-IR spectral energy distribution (SED)
and with the morphology observed in our new mid-IR images.  As a
starting point, we fit the SED of W49/G using the Robitaille et al.\
(2006, 2007) grid of young stellar object (YSO) models and the online
SED fitter.  To construct the mid-IR SEDs shown in
Figure~\ref{fig:SED}, we used the 8--13 $\mu$m spectrum from Gillet et
al.\ (1975), combined with ground-based photometry of source G from
Smith et al.\ (2000) at 12.3, 12.8, and 20.6 $\mu$m, and photometry
obtained at 3.6, 4.5, and 5.8 $\mu$m with the Infrared Array Camera
(IRAC) on the {\it Spitzer Space Telescope} as part of the GLIMPSE
survey (Benjamin et al.\ 2003).  (These additional data were
necessary, as our Gemini data did not have reliable photometric
calibration stars, as noted earlier.)  The ground-based spectral data
at 10~$\mu$m had a large beam (22\arcsec) compared to the photometric
data ($\sim$2\arcsec\ at 3.6, 4.5, 5.8 and 4\arcsec\ at 12.3, 12.8,
20.6), so we scaled the flux in the spectrum to match the imaging
photometry.  It is therefore possible that the silicate absorption
feature at 10 $\mu$m is partially filled-in by silicate emission from
the surrounding heated nebula.  Our Gemini images at 9.7 and
18.5~$\mu$m in Figure~\ref{fig:one} do show a halo that is more
extended than at 11.6~$\mu$m (see also Figs.~\ref{fig:three} and
\ref{fig:four}), so extended silicate emission may contaminate the
larger beam used by Gillet et al.\ (1975).

With this possible contamination in mind, we fit the SED both with and
without the 10 $\mu$m spectral data.  Three representative models are
discussed below and summarized in Table~2.  The SEDs fit to both the
broad-band data points and 10 $\mu$m spectral data (model A in
Fig.~\ref{fig:SED}) correspond to models that consist of a central
source with mass 25--35 $M_{\odot}$, and luminosity 1--2$\times$10$^5$
$L_{\odot}$, a massive accreting envelope with an equivalent accretion
rate $\dot{M}\simeq$10$^{-4}$--10$^{-3}$ $M_{\odot}$ yr$^{-1}$, and
disk mass 0--0.1 $M_{\odot}$ (the mid-IR emission is dominated by
envelope emission, so the SED fits are insensitive to inner disk
mass).  Note that the accretion rate is a model parameter prescribing
the envelope density and mass, and is not a direct measure of the true
accretion rate.  These well-fit SED models (defined as having $\chi^2
- \chi_{best}^2 < 3$, where $\chi_{best}^2$ is the $\chi^2$ value of
the best fit) have viewing angles of 85\arcdeg, and bipolar cavity
opening angles of 11-18\arcdeg.  Fits that allow the silicate
absorption feature to be deeper than indicated by the Gillet et al.\
(1975) spectral data (Model B in Fig.~\ref{fig:SED}), on the other
hand, give similar stellar properties (a central stellar source of
mass 25--35 $M_{\odot}$, and luminosity 2--3$\times$10$^5$
$L_{\odot}$), an envelope accretion rate $\dot{M} \simeq$
5$\times$10$^{-4}$ $M_{\odot}$ yr$^{-1}$, no disk, viewing angles of
30--90 degrees, and very narrow bipolar cavities (2--3\arcdeg).  This
second group of model SEDs has deeper silicate absorption than the
Gillet et al.\ (1975) spectrum, consistent with the extended silicate
emission mentioned above.  However, the very narrow opening angle is
in conflict with the observed mid-IR morphology.

\begin{figure*}\begin{center}
\epsfig{file=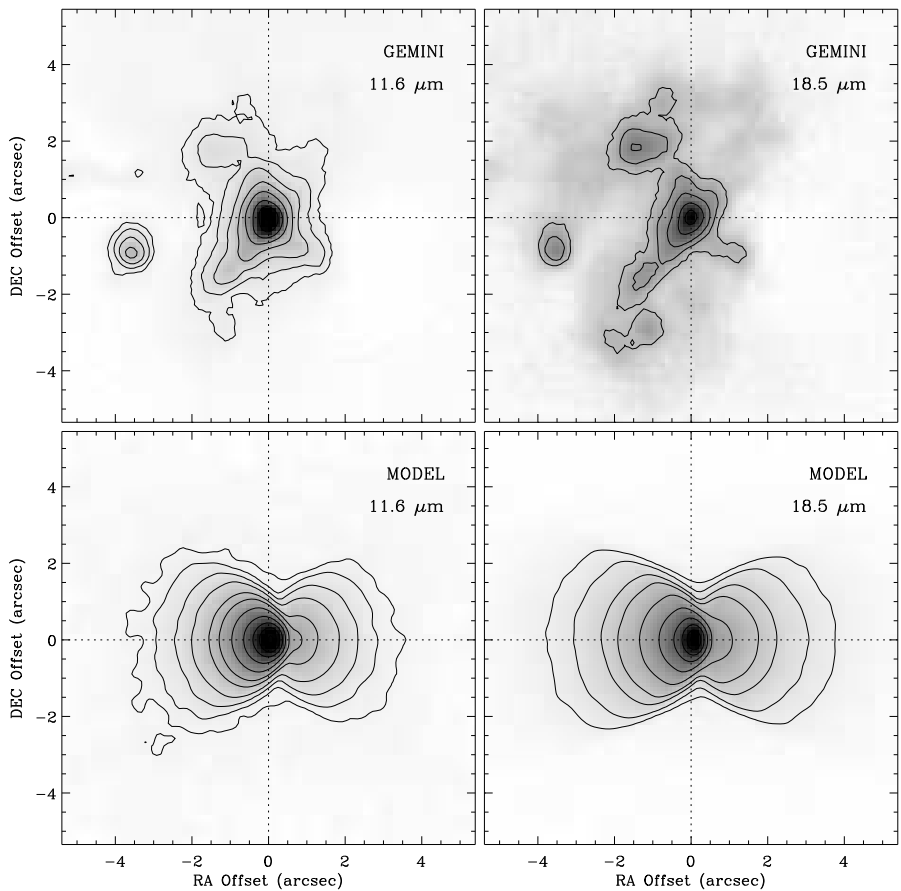,width=4.8in}
\end{center}
\caption{Observed Gemini images at 11.7 and 18.5 $\mu$m (top row)
  compared to model images at the same wavelengths (bottom row).
  These model images correspond to Model C for the SED shown in
  Figure~\ref{fig:SED}.}\label{fig:model}
\end{figure*}

Our high-resolution mid-IR images in Figure~\ref{fig:one} indicate a
relatively large bipolar cavity opening angle ($\sim$30\arcdeg), and
the radio continuum, CO, and water maser data argue for a viewing
angle tilted from edge-on (\S 4.3).  Therefore, we modified the model
parameters accordingly in order to fit both the SEDs and structures in
Gemini images (Fig.~\ref{fig:one}) with a viewing angle of
$\sim$60\arcdeg.  The significantly different result with model C
highlights the value of simultaneously fitting both the SED and the
spatially resolved morphology.  The resulting model IR spectrum is
shown as Model C (black in Fig.~\ref{fig:SED}), and the model images
at 11.6 and 18.5 $\mu$m are compared to the observed Gemini data in
Figure~\ref{fig:model}.  The qualitative agreement is quite good,
especially in the cometary-shaped core.  Note that this model did not
account for the density gradient in the background cloud that was
discussed in \S 4.1 and \S 4.3, so the environment-dependent asymmetry
in the large bipolar cavity (see Figure~\ref{fig:outflow}) is not
addressed by these models.

Model C required a more massive star with a stellar mass of $\sim$45
$M_{\odot}$, luminosity $\sim$3$\times$10$^5$ $L_{\odot}$, an
equivalent envelope accretion rate $\dot{M}$$\simeq$10$^{-3}$
$M_{\odot}$ yr$^{-1}$, and a large inner hole of 1700 AU (0$\farcs$15;
still unresolved in our images), which is larger than the expected
dust sublimation radius.  The more massive star in model C has a
hotter effective temperature and this fits the steeply rising 1--10
$\mu$m spectrum better (see Whitney et al.\ 2004), as does the large
inner hole.  Also, the hotter and more massive star is in better
agreement with the star that is needed to provide the ionizing photon
flux of Source G: De~Pree et al.\ (1997) estimated a source with and
equivalent luminosity and ionizing flux of an O4 V type star to
account for the observed radio continuum emission.  However, models of
the infalling envelope include a parameter, $R_c$, which is the
centrifugal radius inside of which a disk would form.  The SED is well
fit with $R_c \simeq$500--1000 AU, consistent with possible disk
formation.  The proper interpretation of the inner disk hole suggested
by the model fits, within a radius of $\sim$1700 AU, is unclear
because of the uncertainty in fitting a steep SED at short wavelengths
in a source with high extinction.  We regard this inner hole as
tentative and in need of confirmation, but we discuss it as a
possibility in the following section.

\section{DISCUSSION}

The high angular resolution of our new mid-IR images obtained with
Michelle on Gemini South reveal the extended IR morphology of source
G, and show that the concentrated peak of source G:IRS1 is coincident
with a hot molecular core and the origin point of the powerful H$_2$O
maser in W49.  This allows us to bring together several observational
clues that shed light on the nature of the driving source of the
outflow.  The key observational parameters to consider are:

1.  The peak of the mid-IR 8-20 $\mu$m emission is coincident with the
origin point of the H$_2$O maser outflow to within the registration
uncertainty of our data.  Our positional uncertainty of 0$\farcs$07 is
smaller than the size scale of the maser outflow and allows for
meaningful comparison with high-resolution radio continuum data.
Although this association with the water maser was suspected when the
IR source was first discovered by Becklin et al.\ (1973), their
measurements were made using a single element detector with a
7\arcsec\ diameter aperture, leaving much room for ambiguity given the
complex structure of source G.  Our images show that several other
faint sources would have contaminated that aperture as well.  Dreher
et al.\ (1984) compared the IR and radio continuum maps and located
the IR source about 2\arcsec\ ($>$20,000 AU) north of the position we
adopt here, which is clearly in conflict with our new data and would
have precluded it from having any role in the outflow.

2.  The peak of the mid-IR 8-20 $\mu$m emission is {\it not}
coincident with any strong compact radio continuum source.  Although
IRS1 is surrounded by more diffuse radio continuum emission from
source G on large scales (several times larger than the whole maser
outflow), the closest peak of compact radio continuum emission that
signifies an UCHII region is source G2a/b, which is offset more than
0$\farcs$35 or $\sim$4,000 AU to the north. This is beyond our
estimated positional uncertainty.  Our finding that the compact source
G:IRS1 is coincident with the origin of the water masers and not radio
source G2a/b supports the emerging view (De Buizer et al.\ 2005;
Tofani et al.\ 1995) that maser sources tend to be associated with
mid-IR emission rather than radio continuum emission from UCHII
regions.

3.  The peak of the mid-IR 8-20 $\mu$m emission is coincident with a
hot molecular core revealed by tracers such as CH$_3$CN, seen in the
high resolution mm-wavelength maps made by Wilner et al.\ (2001).

4.  This hot core in source G also shows signs of accretion through
inverse P Cygni absorption profiles in CS (Williams et al.\ 2004).
This accretion signature in the hot core is consistent with the high
effective envelope accretion rate in our model fits to the IR SED
(about 10$^{-3}$ $M_{\odot}$ yr$^{-1}$).

5.  The morphology of source G:IRS1 suggests strongly that it is
associated with not only the maser outflow, but a much larger bipolar
outflow and cavity traced by CO emission (Scoville et al.\ 1986),
radio continuum, and X-ray emission.  While the inner source itself
shows no compact radio continum emission, diffuse radio continuum
emission is present on large scales from the bipolar cavity, implying
that some Lyman continuum radiation escapes out the poles.  The large
scale CO outflow and the H$_2$O maser outflow appear to be orthogonal.
Diffuse X-ray emission partly fills the interior cavity of the eastern
lobe (Tsujimoto et al.\ 2006), perhaps marking a reverse shock in the
polar outflow.


This combination of observed properties --- a strong mid-IR source
associated with a hot molecular core and maser emission, but not
strong radio continuum --- is quite similar to the hot core in Orion
that powers the extended BN/KL bipolar molecular outflow.  Even the
apparent geometry is similar, with the water maser outflow oriented
orthogonal to the larger wide-angle bipolar molecular outflow in both
cases. Thus, in many ways, source G in W49 appears to be a more
luminous analog of the OMC-1 outflow, and may provide critical insight
to the formation of very massive stars.  However, there is also an
important difference between these outflows: Unlike Orion, at least
one of W49A/G's outflow cavities is filled with ionized gas seen in
the radio continuum (see Fig.~\ref{fig:outflow}), and appears to have
diffuse X-ray emission.  These ionized cavities span a region roughly
8\arcsec\ or almost 0.5 pc across, larger than the BN/KL outflow.  The
ionizing flux to sustain source G that was deduced by De~Pree et al.\
(1997) requires the equivalent of at least six O6-type stars or a
single O4 star.  Thus, perhaps Source G is simply in a somewhat more
evolved wind-blown cavity stage than BN/KL, more akin to the bipolar
H~{\sc ii} region S106 (Bally et al.\ 1998; Smith et al.\ 2001).
Interestingly, in the case of S~106, the smaller of the two bipolar
lobes is also more deeply embedded in the parent cloud.


Normally, these observations of G:IRS1 as a traditional hot molecular
core that is still in a stage of active envelope accretion would seem
to be at odds with its much larger cavity filled with radio continuum
emission and X-ray emission, because one does not expect a hot core to
produce significant ionizing flux.  The high envelope accretion rate
of 10$^{-3}$ $M_{\odot}$ yr$^{-1}$ that we infer from models of the IR
emission would be sufficient to quench the ionizing flux if the source
were spherical.  One might naturally ask {\it what ionizes the gas in
  the large outflow cavity?}  We note two possible interpretations
that have different implications.

Given its bipolar geometry, G:IRS1 itself may be the source of
ionization for the outflow despite its high envelope accretion rate.
As noted earlier, the equivalent ionizing flux of an O4 star needed to
account for the radio continuum emission is in good agreement with the
massive ($\sim$45 $M_{\odot}$) and luminous ($\sim$3$\times$10$^5$
$L_{\odot}$) star that we infer from models of the IR emission.  Even
though it is a strong mid-IR source inside a hot molecular core, the
embedded massive star may have become hot enough to generate a large
UV luminosity that may be able to escape through lower density regions
in the polar directions.  Our model fits to the SED favor a relatively
hot star.  The required Lyman continuum flux passing through a small
polar ``nozzle'' would exceed the dust Eddington limit (radiation
force on dust grains), but that is part of the basic notion behind
using geometry to allow massive stars to form (e.g., Tan \& McKee
2004).

It is unlikely that stellar radiation alone could have been the agent
responsible for driving the powerful outflow observed in Source G.
From observations of the CO outflow, Scoville et al.\ (1986) derive an
outflow mass of 138 $M_{\odot}$ and a momentum of 3500 $M_{\odot}$ km
s$^{-1}$.  Located within $\pm$4\arcsec\ of the central star, the
outflow dynamical timescale for an average $\sim$25 km s$^{-1}$
outflow speed is roughly 10$^4$ yr.  The momentum supplied by the
stellar radiation field, $L t / c$, during this time falls short by a
factor of $\sim$50 if the stellar luminosity has been constant at the
value of 3$\times$10$^5$ $L_{\odot}$ that we infer from radiative
transfer models.  Similarly, the mass-loss rate required for the
mechanical energy of a stellar wind to power the outflow over the same
time period, assuming a typical O-star wind speed of 1500 km s$^{-1}$,
would need to be roughly 2$\times$10$^{-4}$ $M_{\odot}$ yr$^{-1}$.
This is $\sim$100 times higher than typical main-sequence O star
mass-loss rates in the relevant luminosity range (see Repolust et al.\
2004; Smith 2006).  However, such a high mass-loss rate is still only
20\% of the effective envelope accretion rate that we infer from
models.  Therefore, the likely conclusion is that the large-scale CO
outflow in W49A/G was in fact driven by accretion, at least until very
recently, and that the central star in Source G is therefore an
example of a massive star of $\sim$45 $M_{\odot}$ that is forming by
accretion from a disk.  Our models imply that the inner disk has been
cleared out to radii of 1000--2000 AU, but this depends on the
short-wavelength tail of the IR SED that is also severly affected by
extinction, so we regard this result with some caution. If true, this
must have occured recently compared to the $\sim$10$^4$ yr age of the
large CO outflow.  The H$_2$O maser outflow, on the other hand, is
distributed over a smaller $\sim$1\arcsec\ (0.06 pc) region with
speeds up to $\sim$100 km s$^{-1}$, implying a dynamical age of only
$\sim$500 yr.

This interpretation of W49/G directly supports the picture of single
massive star formation by accretion advocated by Yorke \& Sonnhalter
(2002), Krumholz et al.\ (2005), and others.  Indeed, nearly all the
physical parameters we infer for the central star and outflow of
source G:IRS1 (the stellar mass and luminosity, the bipolar cavity
opening angle, the envelope accretion rate, etc.)  are closely matched
in the numerical simulation studied by Krumholz et al.\ (2005).
Source G:IRS1 may be an excellent observational test case for
developing models of massive star formation that include the
disruptive effects of radiative feedback and stellar winds (e.g.,
Krumholz et al.\ 2005, 2009).  In this context, future observations of
this source with facilities such as {\it ALMA} may be of great
interest in order to confirm the tentative inner disk hole that we
infer.

On the other hand, G:IRS1 is in a crowded and observationally complex
environment, leaving room for other possibilities.  A second
conceivable interpretation is that the outflow cavity from source G is
actually ionized by neighboring O stars that already formed within the
same cloud core.  The best candidate for this type of symbiotic
relationship is the nearby source G2a/b, located just a few thousand
AU to the north in projection (see Fig.~\ref{fig:sketch}).  G2a/b
appears to be a small bipolar UCHII region where the putative polar
axis is aligned with that of G:IRS1 and its large-scale outflow.  This
alignment makes it seem plausible that UV radiation generated by the O
star in G2a/b could escape out the poles to ionize the environment of
source G seen in diffuse radio continuum emission.  This
interpretation highlights the degree to which dense clustered
environments are important in massive star formation where feedback
from neighboring massive stars may join forces in disrupting their
common natal environment.

In any case, G:IRS1 may trace a brief but critical phase in massive
star formation, akin to that of the BN/KL outflow in Orion.  Returning
to the UCHII region lifetime problem mentioned earlier, we note that
among the 40--50 UCHII regions in W49A, G:IRS1 is the only water maser
outflow source and large-scale bipolar outflow.  (Source A, undetected
in the IR, appears to be a fledgeling bipolar cavity and circumstellar
torus or disk; De~Pree et al.\ 1997, 2004.) If the UCHII region
lifetime really is of order 10$^5$ yr, then it is likely that the
maser outflow phase exemplified by Source G is transient, lasting no
longer than a few thousand years.  This is comparable to the dynamical
timescale of the observed outflow in Source G with a characteristic
size scale of a few arcseconds and the fastest expansion speeds of
$\sim$100 km s$^{-1}$ in the H$_2$O maser outflow.  Similarly,
500--1000 yr is the dynamical timescale of the related outflow in the
OMC-1 core (e.g., O'Dell et al.\ 2008).

This returns our attention to the significance of the possible inner
disk hole existing simultaneously with the bipolar outflow.  Namely,
the short timescale involved suggests that this inner disk is being
cleared away at the same time that the H$_2$O maser outflow is driven.
Although direct radiative or stellar wind feedback from the central
star has too little momentum to have powered the large scale CO
outflow over the past 10$^4$ yr, it may drive the much younger maser
outflow.  However, we caution that we only infer the inner disk hole
from the absence of short-wavelength IR emission, so it would be
interesting if future high-resolution techniques could directly
resolve this putative inner disk hole, to catch G:IRS1 in the act of
destroying its own accretion disk and thereby fixing the central
star's final mass.

\smallskip\smallskip\smallskip\smallskip
\noindent {\bf ACKNOWLEDGMENTS}
\smallskip
\scriptsize

We thank C.\ McKee and an anonymous referee for a helpful comments.
N.S.\ was partially supported by NASA through Spitzer grants 1264318
and 30348 administered by JPL.  P.S.C.\ appreciates continued support
from the NSF.  This study has made use of data obtained with {\it
  Spitzer}/IRAC as part of the GLIMPSE Legacy survey.


\end{document}